\documentclass[twocolumn]{pasj01}

\begin{document} 
\Received{2018/07/30}
\Accepted{2018/09/07}

\title{Series-connected array of superconductor-insulator-superconductor junctions in the 100-GHz band heterodyne mixer for FOREST on the Nobeyama 45-m telescope}

\author{Taku \textsc{nakajima},\altaffilmark{1,}$^{*}$
Hirofumi \textsc{inoue},\altaffilmark{2,7}
Yumi \textsc{fujii},\altaffilmark{3,7,}$^{\dag}$
Chieko \textsc{miyazawa},\altaffilmark{4}
Hiroyuki \textsc{iwashita},\altaffilmark{4,5}
Takeshi \textsc{sakai},\altaffilmark{6}
Takashi \textsc{noguchi},\altaffilmark{7,6} and
Akira \textsc{mizuno}\altaffilmark{1}}
\footnotetext[$^\dag$]{Deceased 2015 July 6}

\altaffiltext{1}{Institute for Space-Earth Environmental Research, Nagoya University, Furo-cho, Chikusa-ku, Nagoya, Aichi 464-8601, Japan}
\email{nakajima@isee.nagoya-u.ac.jp}

\altaffiltext{2}{Institute of Astronomy, The University of Tokyo, 2-21-1 Osawa, Mitaka, Tokyo 181-0015, Japan}

\altaffiltext{3}{Department of Astrophysics, Nagoya University, Furo-cho, Chikusa-ku, Nagoya, Aichi 464-8601, Japan}

\altaffiltext{4}{Nobeyama Radio Observatory, National Astronomical Observatory of Japan, 462-2 Nobeyama, Minamimaki, Minamisaku, Nagano 384-1305, Japan}

\altaffiltext{5}{Subaru Telescope, National Astronomical Observatory of Japan, 650 North A'ohoku Place, Hilo, HI 96720, USA}

\altaffiltext{6}{Graduate School of Informatics and Engineering, The University of Electro-Communications, 1-5-1 Chofugaoka, Chofu, Tokyo 182-8585, Japan}

\altaffiltext{7}{Advanced Technology Center, National Astronomical Observatory of Japan, 2-21-1 Osawa, Mitaka, Tokyo 181-8588, Japan}


\maketitle

\begin{abstract}
In this study, we designed and experimentally evaluated a series-connected array of superconductor-insulator-superconductor (SIS) junctions in the 100-GHz band mixer for the multi-beam receiver FOREST on the Nobeyama 45-m millimeter-wave telescope. The construction of the junction chip comprised a waveguide probe antenna, impedance matching circuit, SIS array junction, and choke filter, which were made from a superconducting niobium planar circuit on a quartz substrate. The multi-stage impedance matching circuit between the feed point and the SIS junction was designed as a capacitively loaded transmission line, and it comprised two sections with high ($\sim$90 $\Omega$) and low ($\sim$10 $\Omega$) characteristic impedance transmission lines. The structure of this tuning line was simple and easy to fabricate, and the feed impedance matched with the SIS junction in a wide frequency range. The signal coupling efficiency was more than 92\% and the expected receiver noise temperature was approximately two times the quantum limit for 75--125 GHz based on quantum theory. The array junction devices with 3--6 connected junctions were fabricated and we measured their performance in terms of the receiver noise temperature and gain compression in the laboratory. We successfully developed an array junction device with a receiver noise temperature of $\sim$15--30 K and confirmed that the improvement in the saturation power corresponded to the number of junctions. The newly developed array junction mixer was installed in the FOREST receiver and it successfully detected the $^{12}$CO ($J$ = 1--0) molecular line toward IRC+10216 with the Nobeyama 45-m telescope.
\end{abstract}

\section{Introduction}
The 45-m radio telescope in the Nobeyama Radio Observatory (NRO) is one of the largest millimeter-wave telescopes in the world (e.g., see \cite{mor81}; \cite{aka83}). Many observational studies have been conducted using this telescope, including investigations of star formation, galaxy evolution, and interstellar chemistry. The NRO 45-m telescope has been employed in a wide range of research fields and it is clear that it is an important instrument at present. 

The stable operation of receivers and the development of receivers with new capabilities using advanced technology are important to continue making new scientific discoveries. Several receivers have been developed and operated at the NRO in the millimeter band, especially at $\lambda\sim$3 mm, which is the highest frequency range in the NRO 45-m telescope. For example, \citet{ina87} developed a superconductor-insulator-superconductor (SIS) mixer in 80--120 GHz and constructed a single-beam receiver designated as S100. This SIS mixer was fabricated as an Nb/Al-AlOx/Nb junction and it was constructed with the four junctions connected in series. The mixer mount had two mechanical tuners, which comprised a backshort plunger and a stub-tuner for adjusting the source impedance. Subsequently, \citet{sun93} reported a tunerless mixer mount for the same frequency band. They calculated the embedding impedance as viewed from the SIS junctions and the expected double sideband (DSB) receiver noise temperature based on the theoretical design of the mixer mount. Thus, they successfully developed a low noise SIS receiver ($T_{\rm RX}$(DSB) = 35--70 K) at $f_{\rm LO}$ = 80--120 GHz. Moreover, \citet{sun95} developed a focal plane array receiver designated as S115Q. This receiver was installed with fixed tuned single sideband (SSB) mixer mounts and it had four (2$\times$2) beams. The SSB receiver noise temperature was measured as about 150--200 K at $f_{\rm LO}$ = 102--110 GHz. In 2000, a new multi-beam receiver was developed in the 100-GHz band (25 BEam Array Receiver System; BEARS) and installed in the 45-m telescope \citep{sun00}. This receiver was a focal plane array and it was constructed with 25 (5$\times$5) DSB mixers and feed horns \citep{yam00}. 

Over the next 10 years, several new receivers were developed with novel waveguide-type sideband-separating (2SB) mixer technology and they entered open-use operation in the 45-m telescope. This mixer can separate the radio frequency (RF) signal into an USB (upper sideband) and a LSB (lower sideband) by using a quadrature hybrid in either the signal or local oscillator (LO) path (\cite{ker98}; S. M. X. Claude et al. 2000\footnote{Claude, S. M. X., et al. ALMA Memo 316, 2000\\ $\langle$http://library.nrao.edu/public/memos/alma/main/memo316.pdf$\rangle$.}). Operation of the 2SB mode is required for the receivers on the Atacama Large Millimeter/sub-millimeter Array (ALMA), and many 2SB mixers and receiver systems were developed during this period for millimeter and sub-millimeter telescopes throughout the world. For example, \citet{asa03} and S. Asayama et al. (2004)\footnote{Asayama, S., et al. ALMA Memo 481, 2004\\ $\langle$http://library.nrao.edu/public/memos/alma/main/memo481.pdf$\rangle$.} developed an integrated 2SB mixer in the 100-GHz band, and conducted the first astronomical observations with the waveguide-type 2SB mixer on the prototype ALMA antenna. Moreover, E. F. Lauria et al. (2006)\footnote{Lauria, E. F., et al. ALMA Memo 553, 2006\\ $\langle$http://library.nrao.edu/public/memos/alma/main/memo553.pdf$\rangle$.} and \citet{nak07} reported the first astronomical observations with the SMT (Submillimeter Telescope) and the first large-scale multi-line mapping observations with the 60-cm survey telescope, respectively, with the 2SB mixer in the 200 GHz band. At the NRO, a new single-beam 2SB receiver was developed in the 100-GHz band and installed in the 45-m telescope during 2007. This receiver designated as T100 comprises an OMT (ortho-mode transducer) and two 2SB mixers, which are both based on a waveguide technique. \citet{nak08} reported the first astronomical observation using the waveguide-type dual-polarization 2SB receiver system in the 100-GHz band. 

At that time, the SIS mixer device in the 100-GHz band in the 45-m telescope was a parallel-connected twin-junction (PCTJ) type device. The PCTJ circuit comprises two identical SIS junctions and a stripline inductor to tune out the capacitance of the junctions. The junctions are connected in parallel through the stripline inductor (\cite{nog95}; \cite{shi97}). \citet{asa03} developed a DSB mixer using a PCTJ circuit device and reported that the receiver noise temperature was less than 25 K in the LO frequency range of 95--120 GHz, with an intermediate frequency (IF) range of 4.0--7.5 GHz. Moreover, the SSB receiver noise temperature was measured for the 2SB mixer using this PCTJ device as lower than approximately 50 K over an RF frequency range of 85--120 GHz with 4.0-8.0 GHz IF in the new 2SB receiver system on the 45-m telescope \citep{nak13}. The mixer for the PCTJ device has high performance, as mentioned above, so this design was employed in the mixer for the ALMA band-4 ($f_{\rm RF}$ = 125--163 GHz) and band-8 ($f_{\rm RF}$ = 385--500 GHz) receivers (\cite{sat08}; \cite{asa14}). However, we found that the mixer with the PCTJ device in the 100-GHz band on the 45-m telescope exhibited severe gain saturation when connected to a load of 273 K during chopper-wheel calibration. In fact, we confirmed that the molecular line peak intensity of the standard calibration source for the 45-m telescope, such as the $^{12}$CO ($J$ = 1--0) line in Orion KL, increased by more than 10 \% compared with that for the old array junction device. It is well known that the dynamic range of an SIS mixer depends on $N$ and $f$, where $N$ is the number of series-connected junctions and $f$ is the frequency (\cite{tuc85}), and a theoretical explanation of the gain compression was provided by A. R. Kerr (2002)\footnote{Kerr, A. R. ALMA Memo 401, 2002\\ $\langle$http://library.nrao.edu/public/memos/alma/main/memo401.pdf$\rangle$.}. Therefore, our design of the PCTJ circuit device, with $N$ = 2, exhibited excessively large gain compression at least in the 100-GHz band, and thus it is important to develop a device with a series array of SIS junctions with $N >$2 to improve the linear performance.

In this study, we developed a novel series array of SIS junction device in the 100-GHz band for a new multi-beam receiver system called FOREST (FOur beam REceiver System on the 45-m Telescope). We developed this SIS mixer to meet the following aims.
\begin{enumerate}
\item To design a tuning circuit for a series array of SIS junctions ($N \geq$ 3) in the 100-GHz band ($f_{\rm RF}$ = 80--116 GHz) in order to reduce the gain compression (at least $<$ 10\%) in the black body radiation at room temperature for chopper-wheel calibration.
\item To expand the IF frequency range from 4.0--8.0 GHz to 4.0--12.0 GHz because the IF circuit of FOREST has a wide frequency range compared with the previous receivers in the 45-m telescope. An IF band width of 8 GHz is needed for the simultaneous observation of three carbon monoxide (CO) isotopic molecular lines, which are the most basic and important lines for studies of interstellar molecular gas properties.
\item To develop DSB mixers with the lowest possible receiver noise temperature for the PCTJ circuit device ($\sim$20 K) for the 2SB mixers on FOREST.
\end{enumerate}
In Section 2, we describe the design of each section of the mixer, i.e., the waveguide probe, RF impedance matching circuit, and RF band filter for the IF signal. In Section 3, we present the results measured for the receiver noise temperature and the gain compression for the DSB mixers in the laboratory, as well as test observations using the newly developed mixer in FOREST on the 45-m telescope.

\section{Design of the SIS Mixer}

A SIS junction and RF matching circuit were designed considering a wide RF range (80--116 GHz; fractional bandwidth is 35\%), wide IF range (4--12 GHz), low noise performance ($T_{\rm rx}$(DSB) $\lesssim$20 K), and low gain compression ($<$10\% to 2SB mode with room temperature noise). These goals were determined based on the requirements for open-use of the 45-m telescope.

The SIS mixer design comprised two independent planar circuits. One circuit was constructed with the probe antenna and RF filter to convert from the waveguide mode to the microstrip line (MSL) mode and to reject the RF band for the IF signal, respectively, on a 300 $\mu$m-thick quartz substrate. The other circuit was the impedance matching line between the feed point and the SIS junction on a 300 nm-thick SiO$_{2}$ thin film. The transmission lines in both circuits were fabricated with superconducting niobium (Nb) thin film. The structures of these planar circuits were designed using electromagnetic analysis with HFSS software\footnote{ANSYS HFSS $\langle$https://www.ansys.com/products/electronics/ansys-hfss$\rangle$} and an electrical circuit simulator Genesys\footnote{Keysight Genesys $\langle$https://www.keysight.com/en/pc-1297125/genesys-rf-and-microwave-design-softwares$\rangle$}.

\subsection{Waveguide Probe}

\begin{figure}
 \begin{center}
  \includegraphics[width=8cm]{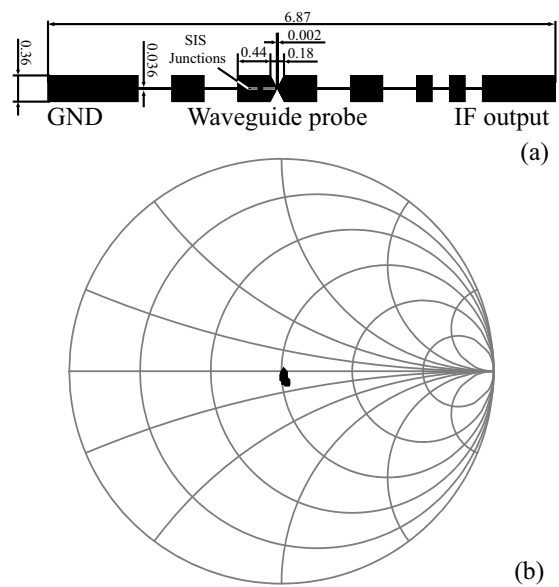} 
 \end{center}
\caption{(a) Pattern of the RF and IF circuit on the SIS mixer chip. The feed point is located in the center of the double-sided bow-tie type probe. (b) Feed point impedance of the waveguide probe on the Smith chart normalized by 70 $\Omega$ and calculated with the HFSS. The frequency range is from 80 to 118 GHz.}\label{fig:chip}
\end{figure}

The construction of the split-block waveguide unit for a mixer, which comprised RF and LO waveguides, a LO coupler, and a channel for the SIS mixer chip, was same as the design employed in a previous study (\cite{asa04}). The RF and LO signals passed through a tapered impedance transformer, which changed the standard W band waveguide (2.54 $\times$ 1.27 mm) into a 1/5 reduced-height waveguide (2.54 $\times$ 0.254 mm) in dimensions, and they were fed to the feed point on the SIS mixer chip. The SIS mixer chip (0.4 $\times$ 7 $\times$ 0.3$^{t}$ mm$^{3}$) was placed into a chip channel in the mixer block and one end of the planar circuit is connected to a 50 $ \Omega$ IF line by a 25 $\mu$m-diameter Al wire to extract the IF output. The channel slot on the other side was filled with indium, which held the SIS chip and connected to electrical ground.

The waveguide probe was a double-sided bow-tie antenna, which was designed by \citet{asa03}, as shown in Figure~\ref{fig:chip}(a). The feed point was located in the center of the bow-tie antenna and it was treated as a lumped-gap-source port in HFSS. The feed point impedance of this waveguide probe is shown in Figure~\ref{fig:chip}(b) and it was designed as 70 $\Omega$. The calculated return loss from the input waveguide to the feed point on the SIS mixer chip was lower than 25 dB in the frequency range of 80--118 GHz. 

\subsection{Series-Connected Array of SIS Junctions}

\begin{figure}
 \begin{center}
  \includegraphics[width=8cm]{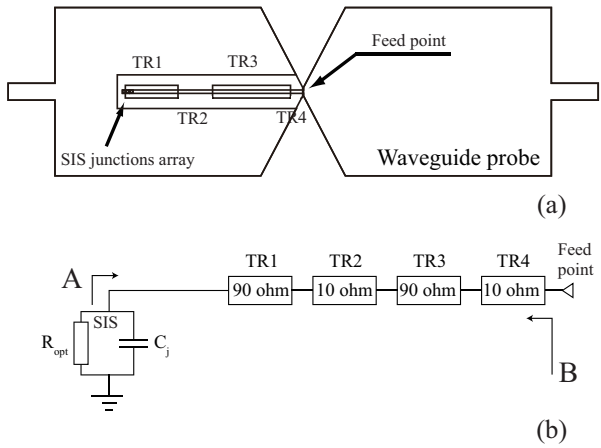} 
 \end{center}
\caption{(a) Design of the impedance transformer from the feed point to the SIS junction array on the bow-tie probe. (b) Model of the RF matching circuit. The circuit comprises four transmission lines. TR1 and TR3 are high-impedance coplanar waveguides, and TR2 and TR4 are low-impedance microstrip lines. The impedances as viewed from the SIS junction (A) and from the feed point (B) are shown in Figure 3.}\label{fig:circuit}
\end{figure}

\begin{figure*}
 \begin{center}
  \includegraphics[width=16cm]{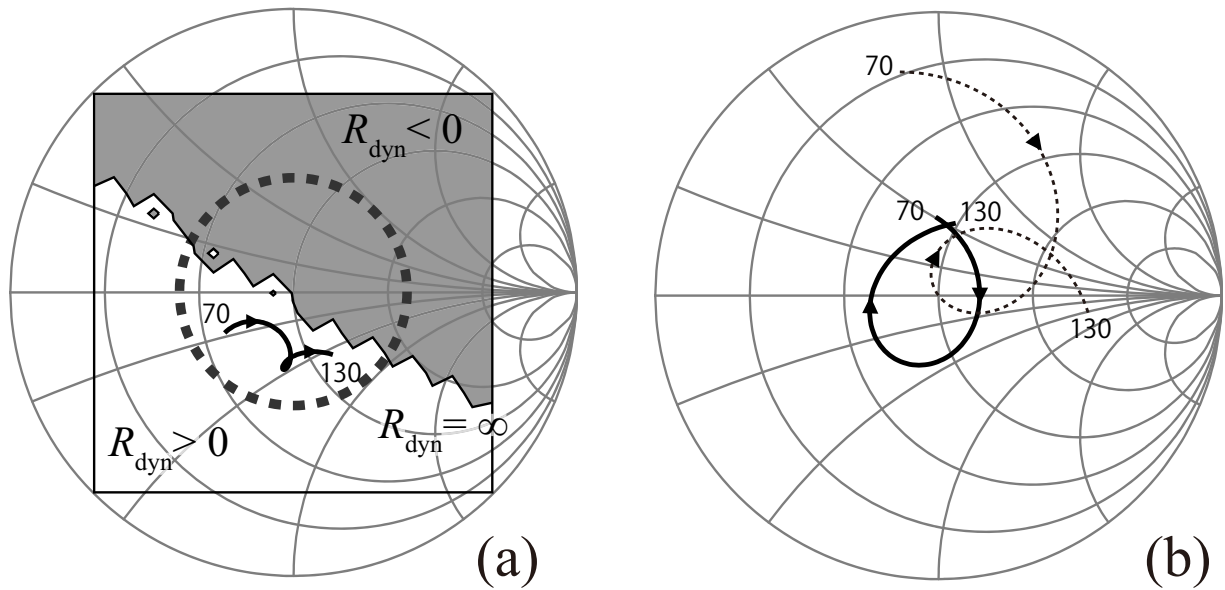} 
 \end{center}
\caption{(a) Embedding impedances as viewed from the SIS junction (at position A in Figure 2(b)) and (b) the impedance through the circuit from the feed point (at position B in Figure 2(b)). The Smith charts for (a) and (b) are normalized by $R_{\rm opt}$ = 25 $\Omega$ and the feed impedance 70 $\Omega$, respectively. Both of the analytical frequency ranges are 70--130 GHz. The dashed circle and the square in (a) represent the magnitude of the reflection coefficient $|\rho|$ $\leq$0.4 and the calculated region of dynamic resistance ($R_{\rm dyn}$), respectively. The gray shaded region at the top right indicates an unstable region where the mixer has negative dynamic resistance. The frequency characteristics of solid and dashed lines are the CLTL circuit in this work and the PCTJ circuit device in the previous work \citep{asa03}, respectively.}\label{fig:smith}
\end{figure*}

\begin{figure*}
 \begin{center}
  \includegraphics[width=16cm]{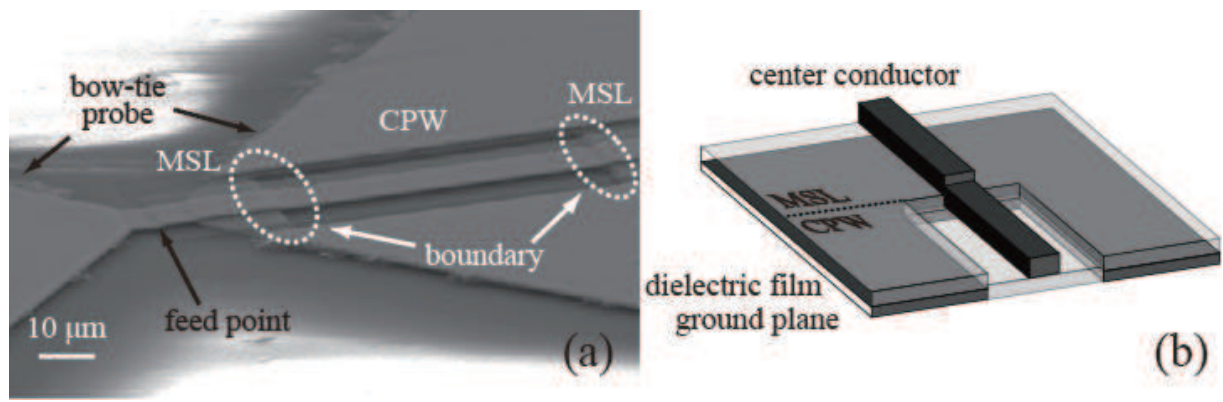} 
 \end{center}
\caption{(a) Scanning electron microscope image of the tuning circuit for the mixer device. The boundary plane between the microstrip line (MSL) and coplanar waveguide (CPW) is the region within the dotted circle. (b) Structure of the capacitively loaded transmission line, which comprises the MSL and CPW where the center conductors and planes are the same, but the center conductor is different at each level.}\label{fig:cpwmsl}
\end{figure*}

According to the theoretical and experimental previous works (e.g. \cite{tuc85} and \cite{ker90}), the RF bandwidth ($\Delta\omega$) was determined from the $\omega R_{\rm n}C_{\rm j}$ product, where $R_{\rm n}$ is the junction normal state resistance and $C_{\rm j}$ is the junction specific capacitance. 1/$\omega R_{\rm n}C_{\rm j} \propto \Delta\omega/\omega_{0} \propto J_{\rm c}/\omega$, so this value was selected 2.8 at the center frequency ($\omega_{0}$) of 97.5 GHz, and thus the junction critical current density ($J_{\rm c}$) was 2.9 kA cm$^{-2}$. The sizes of a single junction and the target $R_{\rm n}$ were to be assumed 1.1 $\times$ 1.1 $\mu$m $^{2}$ and 62.7 $\Omega$, respectively.

In astronomical observations with radio telescopes, the SIS mixers must exhibit good linear performance, i.e., a large dynamic range, between the input noise to the mixer and the output voltage from a detector attached to the receiver. During the calibration of the output signal intensity from the mixer receiver, such as a molecular line spectrum and continuum wave, a black body noise source at room temperature is typically used, which is well known as the chopper-wheel method \citep{uli76}. However, it is known that a single-junction SIS mixer readily saturates in response to a signal with very low ($\sim$pW) input power \citep{fel83}. Thus, a series SIS junction array is important for increasing the dynamic range of the SIS mixer. \citet{tuc85} reported that the saturation power level (P$_{\rm sat}$) can be calculated as follows:
\begin{equation}
P_{\rm sat} \propto \left(\frac{Nh\omega}{e}\right)^{2}\frac{L}{2R_{\rm L}},
\end{equation}
where $N$ is the number of junctions in a series-connected array. Therefore, an array-junction mixer provides an increased dynamic range. 

We designed $N$ = 3, 4, 5, and 6 array junction on a coplanar waveguide (CPW) devices for testing under the assumption of an equivalent single junction \citep{tuc85}. Therefore, the size of each junction was multiplied by the size of a single junction and by $\sqrt[]{\mathstrut N}$. \citet{ton13} noted that it is important to consider the effect of the interconnecting lines between the individual junctions for mixers at a high frequency. The frequency band of our mixer was about 100 GHz and the lengths of the interconnecting lines were 8.5 $\mu$m of CPW, which are smaller than 1/200 of the wavelength ($\sim$1.9 mm), so we considered that the effects of these lines would have been small on the electrical circuit. Therefore, we treated part of the series array junctions as an equivalent single junction. 

\subsection{Impedance Matching Line}

\begin{figure*}
 \begin{center}
  \includegraphics[width=16cm]{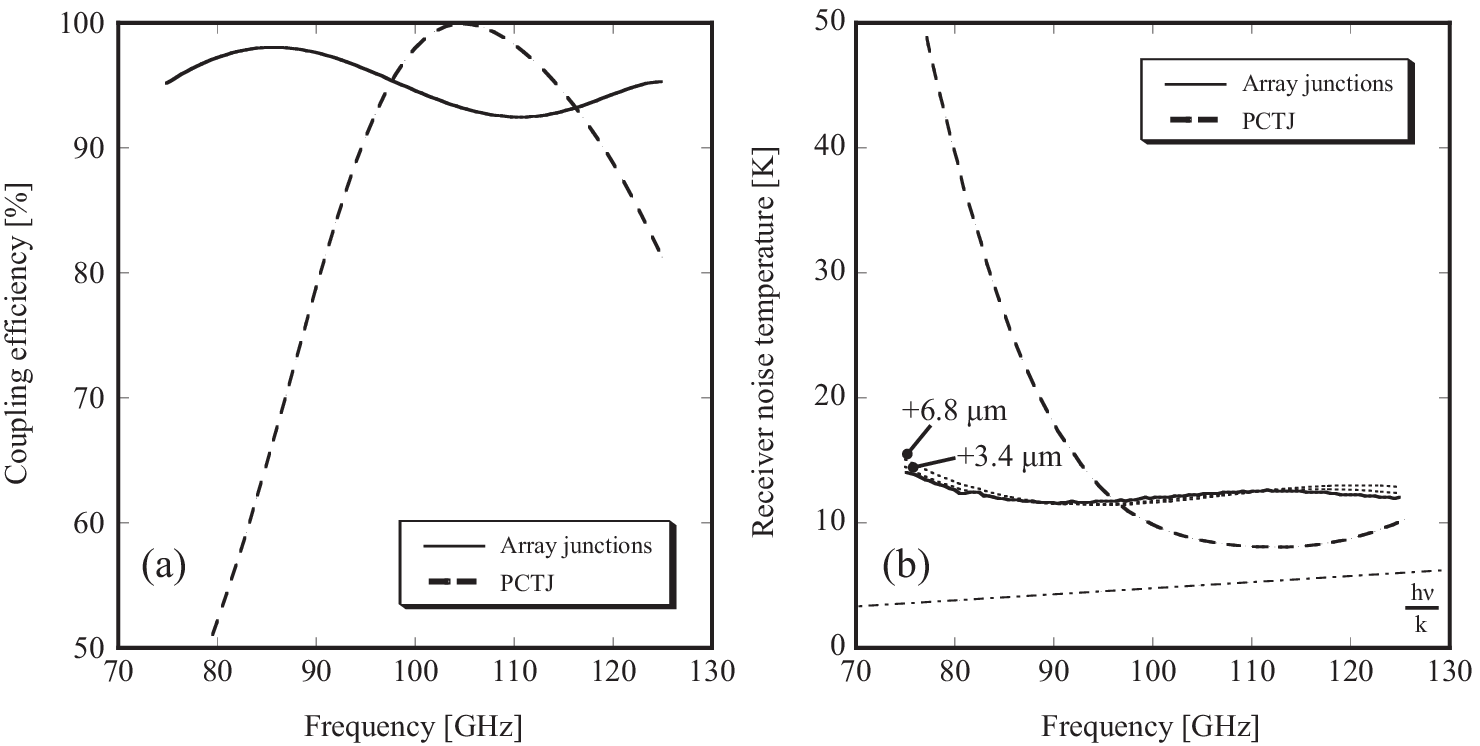} 
 \end{center}
\caption{Simulated coupling efficiency (a) and expected receiver noise temperature (b) based on the designed impedance matching line between the feed point and the SIS junction. Solid and dashed lines represent the array junction device and the PCTJ circuit device, respectively. Dotted lines represent the changes in the noise temperatures due to the effective length extension of the coplanar waveguides for +3.4 $\mu$m (short circuit) and +6.8 $\mu$m (open circuit). The dashed-dotted line in (b) is the quantum limit.}\label{fig:couple}
\end{figure*}

We designed an end-loaded stub type matching circuit for the impedance transformer from the feed point to the SIS junction array. This circuit comprised two sections with high ($\sim$90 $\Omega$) and low ($\sim$10 $\Omega$) characteristic impedance transmission lines, and it was located on the bow-tie probe (Figure~\ref{fig:circuit}(a)). The high and low impedance lines were implemented using a CPW and a superconducting MSL, respectively. Therefore, this circuit can be viewed as a capacitively loaded transmission line (CLTL). This structure and its equivalent circuit are shown in Figure~\ref{fig:circuit}(a) and (b), respectively. As shown in this figure, the structures of SIS junction and TR1 are likely to be a capacitively loaded coplanar waveguide (CLCPW; \cite{ker96}). The impedances as viewed from the SIS junction (position A in Figure~\ref{fig:circuit}(b)) and from the feed point (position B in Figure~\ref{fig:circuit}(b)) are shown in Figure~\ref{fig:smith}. In Figure~\ref{fig:smith}(a), the center of impedance on the Smith chart is where the SIS junction had the lowest noise temperature, but we designed the circuit so the impedance would not enter the unstable region ($R_{\rm dyn} <$ 0 or =$\infty$), as shown by the negative resistance on the I-V curve in the simulation \citep{ino12}. No negative resistance is visible in the bottom left region ($R_{\rm dyn} >$ 0) and the mixer could operate in a stable manner. In addition, for the impedance located in the circle with a radius of 0.4, as represented by the dashed line in Figure~\ref{fig:smith}(a), the SIS mixer exhibited acceptable performance in terms of both the mixer noise temperature and gain \citep{ker95}. In Figure~\ref{fig:smith}(b), the center of impedance on the Smith chart is the feed impedance 70 $\Omega$. Since the CLTL of multi-stage tuning circuit is adopted, the distance from the center is almost similar value over the full RF range on the Smith chart. On the other hand, the PCTJ circuit device show a peaky resonace frequency because it has only one stage impedance transformer \citep{asa03}.

This tuning line comprised two different structure lines, i.e., MSL and CPW, and they were connected inline. Therefore, the boundary plane between the MSL and CPW was a discontinuous structure (Figure~\ref{fig:cpwmsl}). \citet{bei93} suggested that an effective length extension could occur at the edge face of the planar circuit due to a parasitic component. They calculated the effective length extensions at the edge of the CPW as 3.4 $\mu$m for the short circuit and 6.8 $\mu$m for the open circuit. In order to estimate the extension in the electromagnetic analysis, we modeled a two-terminal-pair circuit, where we assumed no physical length and no loss between the MSL and CPW with HFSS. As a result, the Z-matrix for this circuit was inductive and the effective length was longer than the combined physical length of MSL and CPW. The extension was estimated as 4.8 $\mu$m at 100 GHz according to our calculations, where this value was between the lengths estimated for the short and open circuits by \citet{bei93}. Therefore, the lengths of the CPWs in the CLTL circuit used in our mixer device for testing were designed as 3.4 $\mu$m and 6.8 $\mu$m short on both sides of the CPW. However, the expected degradation of the receiver noise temperature due to this effect was approximately as only 1--2 K at both edges of the frequency range covered (Figure~\ref{fig:cpwmsl}(b) because these extension lengths were 4\% of the corresponding wavelength of the CPW at most.

We simulated the performance of the SIS mixer with the CLTL circuit described above. The coupling efficiency of the RF/LO signal through the impedance matching line to the SIS junction is shown in Figure~\ref{fig:couple}(a), and the expected receiver noise temperature based on the calculations is shown in Figure~\ref{fig:couple}(b). When the coupling efficiency is 100\%, all of the input signal is fed to the SIS junction without reflection. We calculated the reflection coefficient ($S_{11}$) between the feed point and the SIS junction based on the calculation of electrical circuit using Genesys software, and estimated the coupling efficiency ($1-|S_{11}|^{2}$). The receiver noise temperature was simulated with the SIS mixer analyzer (SISMA; Shan \& Shi 2007, private communication), which is based on quantum mixing theory \citep{tuc79} with 3-port and 5-port approximations \citep{ton90}. In this calculation, we applied the embedding impedance of the RF circuit calculated with Genesys (Figure~\ref{fig:smith}(a)) and the noise temperature of the IF amplifier was assumed to be 10 K. This calculation excluded the loss of optics, such as the RF window, feed horn, and LO coupler, as well as the IF signal characteristic compared with the actual receiver system. Under these conditions, the expected receiver noise temperature was approximately 10 K, which is twice the quantum limit ($h\nu/k$). Moreover, we confirmed that the array junction receiver covered a wider frequency range than the PCTJ device, i.e., more than 92\% compared with the range of 75--125 GHz.

\subsection{RF filter design for IF characteristic}

\begin{figure}
 \begin{center}
  \includegraphics[width=8cm]{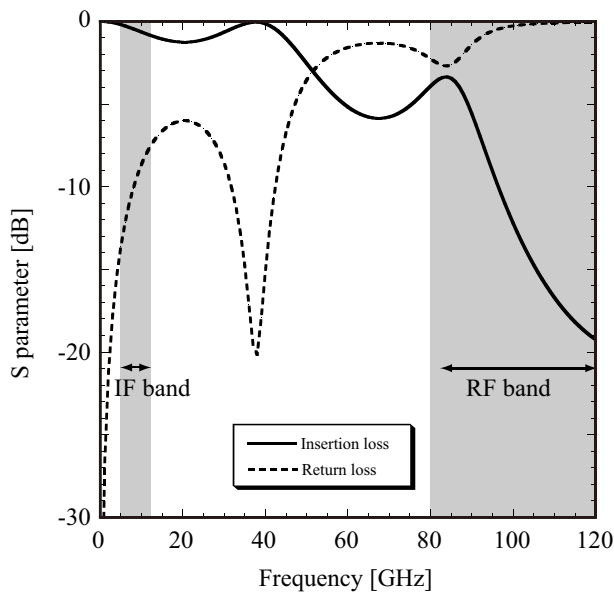} 
 \end{center}
\caption{Simulated return and insertion losses for the choke filter versus frequency. The solid and dashed lines are the insertion and return losses, respectively. The frequency range is 0--120 GHz, which includes the IF and RF frequencies. The shaded region shows the actual IF and RF bandwidth of FOREST.}\label{fig:filter}
\end{figure}

The IF range for FOREST is 4--12 GHz (8 GHz bandwidth) because the molecular lines of $^{12}$CO ($J$ = 1--0; $f_{\rm rest}$ = 115.271 GHz), $^{13}$CO ($J$ = 1--0; $f_{\rm rest}$ = 110.201 GHz), and C$^{18}$O ($J$ = 1--0; $f_{\rm rest}$ = 109.782 GHz) can be obtained simultaneously in a single-sideband. The differences in the remaining rest frequencies between $^{12}$CO and C$^{18}$O is $\sim$5.5 GHz, and the IF bandwidth must be more than this difference. Therefore, it is important to design an RF rejection filter with low insertion loss in this IF range.

The design of the RF filter is shown in Figure~\ref{fig:chip}(a). In order to extract the IF signal from the mixing device, a low-pass filter was needed to transmit the signal for the IF range and to block the RF range from the probe. We designed an RF choke filter based on a previous study by \citet{asa03}. This RF choke filter comprised high and low characteristic impedance 1/4$\lambda$ transmission lines (stepped-impedance low-pass filter), and the RF impedance at the input of the choke filter as viewed from the mixing device was larger by a factor of the squared impedance ratio of the pair. In addition, the impedance section was much shorter than the wavelength of the IF frequencies. A mixer chip with this IF circuit has been used in several receiver systems (e.g., S. Asayama et al. 2004$^{2}$, \cite{nak13}). 

The simulated insertion and return losses of this choke filter calculated with HFSS are shown in Figure~\ref{fig:filter}. The insertion loss of the filter at 12 GHz was approximately --0.8 dB. When we considered a cascade arrangement for the SIS junction, the IF output section including the RF choke filter, and a low noise amplifier, the increase in the receiver noise temperature at 12 GHz was 2 K at most relative to that at 4 GHz under the assumption of 20 K and 0 dB for the noise temperature and conversion gain of the SIS mixer, respectively, and 10 K for the noise temperature of the amplifier. This degradation value did not have a large impact relative to the noise temperature of this receiver system. However, the performance of the RF choke filter at a higher IF frequency ($\gtrsim$12 GHz) was more than -1 dB. Therefore, we designed and tested a new low-pass filter as a hammer type, as suggested by A. Navarrini \& B. Lazareff (2001)\footnote{ Navarrini, A., \& Lazareff, B. ALMA Memo 351, 2001\\ $\langle$http://library.nrao.edu/public/memos/alma/main/memo351.pdf$\rangle$.}, for the ALMA receiver in the 275--370 GHz band. We optimized the size of the hammer pattern to reject the RF frequency band (80--120 GHz) and determined a number for the pattern between the probe and IF output port. We successfully designed large rejection in the RF range ($\lesssim$-30 dB) and a small insertion loss in the IF range from 4 to 20 GHz ($\lesssim$0.5 dB) \citep{koz15}. The details of this filter design will be described in another paper.

\section{Results}

\subsection{Receiver Noise Temperature}

\begin{figure}
 \begin{center}
  \includegraphics[width=8cm]{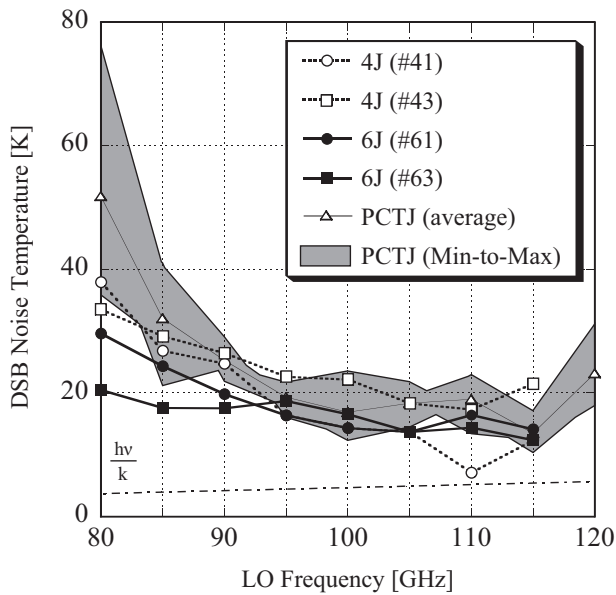} 
 \end{center}
\caption{Noise temperatures measured for the newly developed array junction devices (4J and 6J) and existing PCTJ devices in the DSB mode corresponding to the LO frequency. The measurement system include the SIS mixer as well as the feed horn and IF components. PCTJ devices were measured in 10 mixers independently, and the open triangles and shaded region represent the average values and minimum-to-maximum range, respectively.}\label{fig:noise}
\end{figure}

We measured the noise temperatures for the newly developed array junction devices in the DSB mode using the standard Y-factor method for hot (room temperature) and cold (liquid nitrogen cooled) loads with a power meter. Figure~\ref{fig:noise} shows the results, where we compared the array junction devices (4J and 6J) with existing PCTJ devices. It should be noted that these noise temperatures included the contributions from the SIS mixer but also from the losses due to the feed horn, cooled isolator, and HEMT (high electron mobility transistor) amplifier in the range of 4--8 GHz IF. We successfully developed suitable array junction devices with $\sim$15--30 K for the receiver noise temperature $T_{\rm rx}$, which was almost the same as the performance of the PCTJ device. Moreover, we confirmed that the array junction devices covered a wider frequency range than PCTJ device. The trends in the frequency characteristics of both the array junctions and PCTJ devices were consistent with the expected features based on the circuit design (Figure~\ref{fig:couple}). However, $T_{\rm rx}$ at 12 GHz in the IF frequency range was typically $\sim$1.5 times higher than that at 4 GHz, and this degradation was large compared with the expected loss due to the RF filter as mentioned in 2.4. This increase in the noise temperature at the high IF range was probably due to the conversion gain characteristics of the SIS mixer as well as the loss characteristics of the coaxial cable and the cooled isolator before the HEMT amplifier in this measurement system, rather than the loss due to the RF filter in the mixer chip.

It has been reported that the mixer noise temperature depends on the number of junctions $N$ and it generally increases with $\sqrt[]{\mathstrut N}$ \citep{tuc85}. However, our results for the 4J ($N$ = 4) and 6J ($N$ = 6) devices did not follow this trend as shown in Figure~\ref{fig:noise}. The results were probably affected by the device fabrication quality such as the implementation of $J_{\rm c}$ and the sizes of the junctions, rather than the effect of the number of junctions, because these devices were fabricated in different processes. Furthermore, \citet{cre87} reported that the measured mixer noise temperature increased as $N$ increased and decreased with the value of $\omega R_{\rm n}C_{\rm j}$. According to their results, the effect of $N$ on the increase in the noise temperature in our device ($\omega R_{\rm n}C_{\rm j}$ = 2.8) was probably small.

The extension length effect in the planar circuit was not confirmed adequately in our devices. Our results indicated that this effect possibly had a small impact on the device in the millimeter wavelength range. According to our calculations, the extension lengths were about 4\% at most for the corresponding wavelength of the CPW. Moreover, the expected degradation of the receiver noise temperature due to this effect was approximately only 1--2 K at both edges of the frequency range covered (Figure 4(b)), and it could not be measured by our measurement system. 

\subsection{Gain compression}

As mentioned in the introduction, the gain compression was more than 10\% with PCTJ devices based on the previous 2SB receiver in astronomical observations with the 45-m telescope. Based on the calculations reported by A. R. Kerr (2002)$^{4}$, 10--20 \% gain compression for the $N$ = 2 mixer under operation in the 2SB mode in the 100-GHz band corresponds to a mixer gain of 0--4 dB. This value is fairly consistent with the maximum mixer gain calculated for the PCTJ type receiver at around 110 GHz according to SISMA (Figure~\ref{fig:couple}). The gain compression for the array junction receiver was expected to improve and it was estimated at $\sim\frac{1}{N^{2}}$ times smaller than that with the PCTJ type receiver if they have the same conversion gain.

We measured the maximum gain compression values for the $N$ = 3 and $N$ = 5 mixers at a load of 273 K in the laboratory and the values obtained were 21.2$\pm$1.2\% and 10.6$\pm$0.7\%, respectively. Therefore, the equivalent gain compression values in the 2SB mode were calculated as about 12\% for $N$ = 3 and 6\% for $N$ = 5 mixers. The value was larger than the target performance for $N$ = 3, but that for $N$ =5 was suitable to meet the requirements for open-use with FOREST on the 45-m telescope. The detailed measurements of the gain compression for our mixer, including the measurement system and method, will be presented in a separate study (T. Nakajima et al. in prep.).

\subsection{First Light of Molecular Line on the Telescope}

\begin{figure}
 \begin{center}
  \includegraphics[width=8cm]{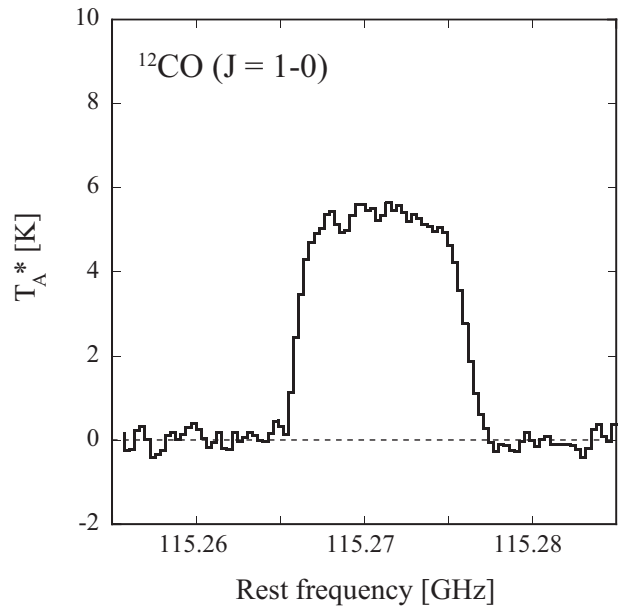} 
 \end{center}
\caption{First $^{12}$CO ($J$ = 1--0) molecular spectrum toward a carbon star obtained with the newly developed array junction mixer on the FOREST receiver.}\label{fig:12co}
\end{figure}

After evaluating the receiver noise temperature and its frequency characteristics with the array junction devices in the laboratory, we developed eight 2SB mixers for the FOREST receiver with four beams, where they could simultaneously obtain the dual polarization signals as well as the upper and lower sidebands in each polarization. Details of the receiver system for FOREST were described previously by \citet{nak12}, \citet{mur13}, and \citet{min16}. The first astronomical signal was obtained from the $^{12}$CO ($J$ = 1--0) spectrum in the USB toward the well-known carbon star IRC+10216 on May 19, 2011 with FOREST using the array junction mixer. Figure~\ref{fig:12co} shows the spectrum received on beam-1 at the center point of this object. Test observations of simultaneous $^{12}$CO ($J$ = 1--0), $^{13}$CO ($J$ = 1--0), and C$^{18}$O ($J$ = 1--0) spectra obtained with the four-beam 2SB mixer receiver toward the Orion KL molecular cloud were presented by \citet{min16}. At present, FOREST is using one of these receivers for open-use on the 45-m telescope and the system noise temperature including the atmosphere is typically 150--300 K. 

\section{Summary}

In this study, we developed a series-connected array of SIS junction device in the 100-GHz band DSB mixer and it was installed in the multi-beam receiver FOREST on the NRO 45-m telescope. An array junction mixer ($N \geq$ 3) was designed in order to reduce the gain compression under hot load temperatures compared with the existing PCTJ type mixer. 

The multi-stage impedance transformer from the feed point to the junction array employed a CLTL, which comprised two sections with high ($\sim$90 $\Omega$) and low ($\sim$10 $\Omega$) characteristic impedance transmission lines. The structure of this tuning line is simple and easy to fabricate, and the RF impedance matched in a wide frequency range. The RF/LO coupling efficiency was more than 92\% and the expected receiver noise temperature was approximately two times the quantum limit, at least from 75 GHz to 125 GHz. 

We fabricated series-connected array devices where $N$ = 3--6 and measured their performance in the laboratory. As a result, we successfully developed suitable array junction devices with $\sim$15--30 K of $T_{\rm rx}$ (DSB) in the range of 80--120 GHz. The minimum value was fairly similar to the performance of the PCTJ device, but the RF frequency range was significantly wider than that for the existing device. We considered the effective length extension at the edge of planar circuit, but the impact of this effect on the noise performance was not clearly confirmed. Our results indicated that this effect possibly had a small impact on the device in the millimeter wavelength range, or it may have been too small to measure with our measurement system.

The gain compression values were measured as 21.2$\pm$1.2\% and 10.6$\pm$0.7\% for the $N$ = 3 and $N$ = 5 mixers, respectively, at a load of 273 K in the DSB mode. Thus, the equivalent gain compression with the $N$ = 5 device using a 2SB mixer was expected to be approximately 6\%, which is sufficient to meet the requirements for open-use with FOREST on the 45-m telescope.

Finally, the newly developed array junction mixer was installed in the FOREST receiver and we successfully obtained the $^{12}$CO ($J$ = 1--0) molecular line toward IRC+10216 with the NRO 45-m telescope.

\begin{ack}
The authors would like to thank Tetsuya Katase and Makoto Koyano at Osaka Prefecture University, and Makio Ito, Naoki Akiyama, Kouta Zengyo, and Masahiro Suzuki at Nagoya University for their contributions to measuring the performance of the SIS mixer. We also wish to thank Tomonori Tamura and Shohei Ezaki for the fabricating the SIS device at the Advanced Technology Center (ATC), National Astronomical Observatory of Japan, and Chihaya Kato for provide the SEM image of the SIS device. We are also grateful to Shin'ichiro Asayama, Jun Maekawa, Nario Kuno, Ryohei Kawabe, and the entire staff of the Nobeyama Radio Observatory for their useful discussions and support.
\end{ack}

\end{document}